# Black hole immunity theorem and dark matter

L. K. Chavda[1] and Abhijit L. Chavda [2]
[1, 2] Chavda Research Institute
49 Gandhi Society
City Light Road
Via Parle Point
Surat 395007
Gujarat, India.
E-mail: [1] holeum@yahoo.com, [2] al_chavda@rediffmail.com

ABSTRACT

We prove that the Primordial Black Holes (PBHs) in the early universe did not decay until the gravity decoupled from the other interactions. Rabinowitz' two objections to the formation of Holeums are shown to be inapplicable. We show that PBHs having masses between $8 \cdot 10^{18}$ GeV and $10^{19}$ GeV formed gravitational bound states called Holeums when the temperature of the universe was between $10^{30}$ K and $10^{29}$ K. We prove that the black holes in a fire ball produced in an accelerator such as the LHC, modified into an inertial confinement type set up, will not emit Hawking Radiation as long as the fire ball is maintained in thermal equilibrium or is allowed to expand adiabatically.

It is generally believed that the submicroscopic Primordial Black Holes (PBH) produced in the Big Bang got evaporated away instantly at their births. This belief arises from the Hawking theorem [1]. The latter is based on the phenomenon of vacuum fluctuations which can be accounted for only in the second-quantized field theory. But during the past eighty years we have not been able to arrive at a consistent quantum field theory of gravity. Therefore the theorem is semi-classical. There is no direct or indirect experimental verification of the Hawking Radiation (HR) for the past thirty years since it was predicted. Belinsky [2] believes that the HR does not exist. Helfer [3] questions it but prefers to await the arrival of the Quantum Gravity (QG) for the final word. Balbinot [4] has shown that highly charged black holes do not emit HR. In reference [5] it was pointed out that HR is valid only for isolated black holes in field-free space. Here we do not question the HR or the Hawking theorem. But we will show that the hypothesis of instant evaporation is a misconception. We will also show that (a) PBHs did not decay prior to the decoupling of gravity from the other interactions in the early universe. (b) Atoms of PBHs, called Holeums, formed during his period. (c) HR can be controlled in an inertial confinement type accelerator set-up. To this end we note that the life-time of a black hole of mass m is given by [1]

$$T = 6.02 \cdot 10^{-18} \, m^3 \, s \qquad (1)$$

in mks units. From this formula we calculate the life-time of a black hole of mass $m_P = (\hbar c/G)^{1/2} = 1.22 \cdot 10^{19}$ GeV, the Planck mass. It is given by

$$\tau_P = 6.20 \cdot 10^{-41} \, s \qquad (2)$$

And the Planck time, characteristic of the Quantum Gravity Regime (QGR), is given by

$$t_P = (\hbar G/c^5)^{1/2} = 5.39 \cdot 10^{-44} \, s \qquad (3)$$

On the other hand, the age of the universe at the Planck epoch is given by

$$\tau_u = 1.61 \cdot 10^{-44} \, s \qquad (4)$$

A comparison of $\tau_P$ with $\tau_u$ shows that the PBH of Planck mass has a life time nearly four thousand times greater than the age of the universe at the Planck epoch. A comparison of $\tau_P$ and $t_P$ shows that such PBHs had enough time to interact thousands of times with one another. Using equation (1) one can show that the Hawking formula does not imply instant evaporation for PBHs of mass greater than about $0.1 m_P$. This directly refutes the "instant evaporation" hypothesis. Secondly since the vacuum fluctuations in the early universe produced a vast quantity of submicroscopic PBHs, it would be frivolous of nature to produce such a vast quantity merely to fritter it away. Thirdly the instant evaporation hypothesis violates the Principle of Nuclear Democracy (PND), namely, all primordial particles like quarks, leptons, PBHs etc., are on an equal footing. None is more important than the others. Whereas the quarks and leptons have played vital roles in the constitution and evolution of the universe, the instant evaporation hypothesis will leave the PBHs with no role at all. This violates the PND. Fourthly Hawking's theorem strictly applies to an isolated black hole in a field-free space. And there were no isolated PBHs in the early universe nor were they in field-free space. Thus, the hypothesis of instant evaporation is a misconception. It is obvious that in the presence of other black holes there will be a mutual give-and-take of HR and the black holes may live longer or may not decay at all. In the latter context we consider two cases of interest in the laboratory and cosmology, namely black holes in thermodynamic equilibrium and those in the expanding universe, respectively. First we consider the latter. The expansion of the universe is generally assumed to be adiabatic. That is, it is reversible. The entropy of the system remains constant during the expansion. Although this is an excellent approximation due to the immense quantity of entropy contained in the cosmic microwave back ground, it is strictly not true. If it were, all the structure we see in the universe today would not be there. Besides, irreversible microscopic processes are always present. Therefore we turn to the uniform cosmological model [6]. In this model the primordial brew containing the quarks, leptons, PBHs, etc. expands adiathermally. That is, there is no net flow of heat through any surface in it. If it were also reversible, then the entropy would remain constant. And we would have an adiabatic expansion. But here we assume the uniform cosmological model with an adiathermal expansion and not an adiabatic one. We prove the following theorem having important implications for the physics of the early universe.

Black Hole Immunity Theorem (BHIT): PBHs do not decay if the primordial brew containing them expands adiathermally.

Proof: A black hole has an event horizon. And the requirement of "no net flow of heat through any surface in it" cannot be satisfied across an event horizon. The presence of the black holes in the primordial brew poses the same kind of problem as the presence of singularities in an analytic function in a given region. The presence of singularities makes the region a multiply- connected one. By cutting out the singularities and connecting them by arbitrarily close cross- cuts we obtain a singly-connected region in which the function is analytic. Similarly here, too, we can look upon the region outside the event horizons as a simply connected one. Now if take an arbitrary surface *lying completely within the simply-connected region*, then the requirement of "no net flow of heat" through it will be satisfied if every PBH uniformly emits as much HR as it absorbs. That black holes emit black body radiation is well- known. Since the rates of emission and absorption of HR are the same, therefore the black holes do not decay. This proves the BHIT.

Implicit in the above theorem is the requirement that the rate of collisions among the constituents of the primordial brew, including the PBHs, exceed the rate of expansion of the universe. This guarantees that there is a sufficient number of collisions to ensure the condition of no net flow of heat. This requirement is, indeed satisfied by the PBHs in the brew until the gravity decouples from the other interactions at the unification temperature $T_u$ where $kT_u = 10^{29}$ K ($10^{16}$ GeV). Here k is the Boltzmann constant. Below $T_u$, the gravity becomes the weakest of the four fundamental interactions of nature. And the PBHs decouple from the brew and would get destroyed due to their HR. But above $T_u$, the rate of gravitational interactions among the PBHs exceeds that of the expansion of the universe. This makes them the bona-fide members of the primordial brew that satisfies the requirement of no net flow of heat. Hence the BHIT applies to them above $T_u$. Thus we have arrived at the following corollary.

Corollary1: The PBHs in the early universe did not decay until the gravitational interaction decoupled from the other interactions at the unification temperature $T_u$.

Although so far we have considered an adiathermal expansion the following corollary may be useful in the laboratory conditions in the near future.

Corollary2 : Black holes in thermodynamic equilibrium or in adiabatically expanding Fire Ball (FB) do not decay.

Proof: Thermodynamic equilibrium and an adiabatic expansion both satisfy the requirement of no net flow of heat through any surface. Hence both meet the requirement of BHIT. Thus, the corollary2 follows from the former. This corollary gives us a handle on the control of HR as discussed below.

Now we consider an application of the BHIT and its corollary1. It concerns the formation of atoms of PBHs called Holeums which are dark matter candidates. Because of the vast quantity of the PBHs produced in the vacuum fluctuations in the early universe, the Holeums are likely to form an important component of dark matter in the present day universe. First we derive the necessary condition for the formation of a Holeum. Let B be the binding energy of a Holeum and let $<K(T)>$ be the average kinetic energy of the PBHs at temperature T of the formation of a Holeum. Then the necessary condition is as follows.

$$B > <K(T)> \tag{5}$$

Now the corollary1 tells us that the PBHs are immune to decay above $T_u$. Thus

$$T > T_u \tag{6}$$

This is a sufficient condition. Since the PBHs do not decay above $T_u$ and since gravity is an attractive interaction (except during the inflation when the vacuum energy dominated), therefore the Holeum formation may occur. Now the binding energy of a Holeum is given by [5]

$$B = (\mu c^2/2)\, \alpha_g^{\,2} \tag{7}$$

where $\mu$ is the reduced mass of two black holes of masses $m_1$ and $m_2$ and we take the equal mass case for simplicity and

$$\alpha_g = (m/m_P)^2 \tag{8}$$

is the gravitational coupling constant and m is the mass of the two identical constituent PBHs. Now the condition for a stable Holeum is [5]

$$0 < m < m_c = (\pi^{1/2}/2)\, m_P = 1.0821 \cdot 10^{19}\, \text{GeV} \tag{9}$$

Now from Table 1 we see that

$$B > 3.68 \cdot 10^{17}\, \text{GeV} \tag{10}$$

when

$$8 \cdot 10^{18}\, \text{GeV} < m < 1.0821 \cdot 10^{19}\, \text{GeV} \tag{11}$$

Thus, if we take

$$<K(T)> \, < 3.68 \cdot 10^{17}\, \text{GeV} \tag{12}$$

and $T > T_u$, then the equation (10) will be satisfied. And that, in turn, will satisfy the necessary condition, equation (5). Now in quantitative cosmology adiabatic expansion is assumed because it is an excellent approximation. Therefore we take

$$<K(T)> = 3kT \tag{13}$$

Then substitution into equation (12) gives us the following range of temperatures for Holeum formation.

$$10^{16}\, \text{GeV} < kT < 10^{17}\, \text{GeV} \tag{14}$$

This is an indicative solution. In general, there is an infinite number of solutions depending upon what value of B one chooses above $10^{16}$ GeV. But all of them favour the Holeum formation close to the maximum value $m_c$ and in a temperature range above $T_u$.

Now we must consider the dynamics of Holeum formation. It is governed by the Saha equation whose derivation follows the recombination theory [6]. We first summarize the main points of similarity. Recombination is the name of the process by which the electrons and the protons in the primordial brew form hydrogen atoms. This occurs at about kT = 0.1 eV to 0.3 eV or T around 1000 K. The binding energy of the hydrogen atom is 13.6 eV. Thus, B/kT is of the order 100 or less. From equations (10) and (12) and the Table 1 we see that B for a Holeum can be as large as $10^{18}$ GeV and the temperature can be as low as $10^{16}$ GeV. Therefore here, too, B/kT is in a similar range. (2) During the recombination era the electromagnetic interactions among the electrons and the protons in the primordial brew are so fast that the expansion of the universe is initially regarded as quasi-static for the purpose of deriving the Saha equation [6]. Now above $T_u$ all the four fundamental interactions of nature are equally strong and their interaction rates, too, are much faster than the rate of expansion of the universe. Thus here, too, one may regard the expansion as quasi-static. (3) Moreover, the PBHs are immune to decay above $T_u$ and the gravity is assumed to be an attractive interaction (except during the inflation when the vacuum energy dominated). This completes the similarity with the recombination era. In fact, since the Holeum is the gravitational avatar of the hydrogen atom, we may call the era starting with the Big Bang and ending with the decoupling of gravity the "Gravitational Recombination Era". Let n be the number density of the PBHs and $n_H$ the number density of the Holeums. Then following [6] we write the Saha equation for this case as

$$n^2/n_H = \lambda^{-3} (kT/2\pi mc^2)^{3/2} \exp(-B/kT) \qquad (15)$$

where

$$\lambda = \hbar/mc \qquad (16)$$

is the Compton wave-length of the PBHs. From the Table 1 we see that B is a very sensitive function of m. For example, when the value of m falls by one order of magnitude, i.e. from $10^{19}$ GeV to $10^{18}$ GeV, B falls by five orders of magnitude. This means that, in the temperature range T > $T_u$, only the PBHs havng masses close to, but less than, $m_c$ would be able to form stable Holeums. But since the PBHs are immune to decay in this range, the other PBHs will have only one option open to them, namely, coalescence. They will go on forming bigger and bigger black holes by coalescence until the gravity decouples at the temperature $T_u$. Here we treat the latter as an empirical parameter with the value $kT_u = 5 \cdot 10^{15}$ GeV. Then eqution (15) gives $n^2/n_H = 2 \cdot 10^{-6}$ cm$^{-3}$. If we take a slightly smaller value kT = $4 \cdot 10^{15}$ GeV, then we get $n^2/n_H = 5.5 \cdot 10^{-31}$ cm$^{-3}$. This confirms the completion of Holeum formation at the temperature around $T_u$. The gravity decouples below $T_u$. The Holeums freeze-out and the black holes begin to decay. If some of the latter are massive enough to last till today and beyond, then one can say that the early universe fulfilled its obligations to the PND during the gravitational recombination era by creating stable Holeums and long-lived macro black holes.

Thus the BHIT and the Saha equation provide a complete theory of formation of Holeum. But the central question is whether the unstable PBHs can form stable Holeums. The BHIT itself tells us that the black holes do not decay at all under certain circumstances. This fact, together with the following considerations, gives us a strong signal that stable Holeums may exist. A century ago Bohr argued that since the matter waves associated with an electron in a hydrogen atom formed standing wave patterns consistent with the quantization of the orbital angular momentum, therefore the electron would not emit the electromagnetic radiation while moving in such orbits. Hence the hydrogen atom would be stable. We can argue similarly because a Holeum is a gravitational analogue of the hydrogen atom. As already mentioned above, the PND requires the existence of stable Holeums. Now Gell-Mann's inevitability principle says that, if something is not forbidden by the laws and principles of physics, then it is inevitable. Now gravity is an attractive interaction (except when the vacuum energy dominates) and the universe abounds in the macroscopic gravitational bound states like the planetary systems, the binary systems and the galaxies. Therefore there is nothing against the formation of microscopic gravitational bound states. And the gravity is so weak that microscopic gravitational bound states must involve the black holes. Thus, the formation of Holeums seems to be inevitable. Besides, in nature we do have unstable particles that form stable bound states. For example, the unstable neutron forms stable nuclei. Another analogy is that semi-free quarks participating in a hard-scattering Drell-Yan process emit a jet of hadrons but the quarks bound in hadrons never do so. The quarks are the primordial particles that exist only in bound states. If the stable Holeums exist then PBHs would join the ranks of their primordial partners the quarks in that both would exist only in bound states.

Recently Rabinowitz [7] has shown that the condition
$$kT \gg mc^2 \qquad (17)$$
which we assumed in reference [5] leads to kinetic energies of the PBHs in excess of their binding energies. That is, our equation (5) is violated. This is a very important point and we concede it. But we would like to point out that the equation (17) represents a sufficient condition and not a necessary one. It implies a relativistic gas of PBHs with enormous number density [5]. We assumed it to create as favourable conditions as possible for the creation of the Holeums. But now in the light of our BHIT and corollary1, we see that it is not needed for Holeum formation. In a nut-shell, by BHIT and the corollary1 the PBHs are immune to decay above $T_u$. It does not matter whether they are relativistic or non-relativistic. Thus, we drop the equation (17) and the Rabinowitz objection disappears.

His second objection is based on the violation of the condition of validity of Newtonian Gravity (NG). He argues that $n \geq 10$ and $r > 10R_H$ are the reasonable criteria for the validity of NG. Here $R_H$ is the Schwarzschild radius of the black hole. Consider the following counter example. We know that the Bohr quantization is a semi-classical result valid for $n \gg 1$, say, $n \geq 10$. We apply this criterion to quantize the Harmonic Oscillator and get the well-known equally-spaced energy levels. But for $n = 1$ the zero-point energy is missing. This is as expected and it vindicates the criterion $n \geq 10$. But if we apply Bohr quantization to the $r^{-1}$ potential (hydrogen atom), we get exact energy eigen-values even for $n = 1$. In other words, we may get exact results even if the criterion $n \geq 10$ is violated. This shows that the criterion is a sufficient condition and not a necessary one. In fact, the $r^{-1}$ potential gives exact results because of an accidental symmetry. But that is precisely our strength. The Holeum is a gravitational avatar of the hydrogen atom. Both follow from the $r^{-1}$ potential and are expected to give exact results for all n. In reference [5] we have meticulously discussed the validity of our results at length. In particular, we have shown that as far as an order of magnitude values of the bound state parameters are concerned, the actual behaviour of the potential (even an infinite potential) within the bound state as well as the finite size of the constituents of the bound state are of no consequence and that only the correct asymptotic behaviour of the potential suffices [5]. Thus, Rabinowitz' second objection is not applicable to the $r^{-1}$ potential of the Holeum.

Now we consider an application of corollary2. The latter shows us how to control the HR from the black holes. But the condition $kT > 10^{16}$ GeV seems to put it beyond the reach of any foreseeable accelerator. However, in a four-plus-two dimensional universe, with the two extra dimensions compactified at the millimeter scale, the Planck energy scale, representing the gateway to the QGR, comes down from $10^{19}$ GeV to 1 TeV and with it $T_u$ also comes down below 1 TeV [8]. Several estimates suggest a copious production of black holes at the Large Hadron Collider (LHC), a 15 TeV accelerator to be operational in 2007 [9]. Now consider an inertial confinement type modification of such an accelerator [10]. Here the main proton-antiproton beams will produce a FB at the temperature of 30 TeV (= 3·$10^{17}$ K) at the centre of a sphere of a suitable radius. The FB will start expanding and cooling. However, 30 TeV energy is being continuously pumped into the centre of the FB. The expanding FB will be compressed uniformly by auxiliary proton-antiproton and/or laser beams from the periphery of the sphere. If the energy of the compressing beams is tunable then we can control the temperature of the FB at will. If we maintain it at a constant value above 1 TeV, the threshold to the QGR in the four-plus-two dimensional universe, then by the above-mentioned theorem and the corollaries the black holes in the FB will not decay. This is so because from 15 TeV to 1TeV we are in the QGR and the gravity has not yet decoupled. Therefore the BHIT and corollaries apply and there will be no HR from the black holes in the FB. But when the condition is broken and the temperature falls below 1 TeV, the HR will be released. Similarly the tunability allows us to let the FB expand adiabatically also. And by corollary2 the black holes in the FB will not emit the HR as long as the temperature is above 1TeV. In this way we can control the HR. The controlled release of HR can have far-reaching consequences. It could be the third mile- stone in energy production after fission and fusion.

Table Caption
Table 1. Binding energies of Holeums as a function of m, equation (7).

Table 1

| m (GeV) | B (GeV) |
|---|---|
| $1 \cdot 10^{18}$ | $1.1244 \cdot 10^{13}$ |
| $2 \cdot 10^{18}$ | $3.5982 \cdot 10^{14}$ |
| $3 \cdot 10^{18}$ | $2.7324 \cdot 10^{15}$ |
| $4 \cdot 10^{18}$ | $1.1514 \cdot 10^{16}$ |
| $5 \cdot 10^{18}$ | $3.5139 \cdot 10^{16}$ |
| $6 \cdot 10^{18}$ | $8.7436 \cdot 10^{16}$ |
| $7 \cdot 10^{18}$ | $1.8898 \cdot 10^{17}$ |
| $8 \cdot 10^{18}$ | $3.6846 \cdot 10^{17}$ |
| $9 \cdot 10^{18}$ | $6.6396 \cdot 10^{17}$ |
| $1 \cdot 10^{19}$ | $1.1244 \cdot 10^{18}$ |